\documentclass{ws-procs9x6}

\begin{document}

\title{
Dynamical baryon resonances from chiral unitarity}

%

\author{A.~RAMOS}
\address{
Departament d'Estructura i
Constituents de la Mat\`eria, Universitat de
Barcelona,  E-08028 Barcelona, Spain}

\author{C.~BENNHOLD}
\address{ Center for Nuclear Studies, Department of Physics,
 The George Washington University, Washington D.C. 20052
 }%
 \author{A.~HOSAKA, T.~HYODO}
\address{ Research Center for Nuclear Physics, Osaka University,
Ibaraki, Osaka 567-0047, Japan
 }%
 \author{D.~JIDO}
\address{ECT$^*$, Villa Tambosi, Strada delle Tabarelle 286, I-38050 Villazzano, Italy
 }%
 \author{U.-G.~MEISSNER}
\address{ HISKP, University of Bonn, Nu\ss alle 14-16, D-53115 Bonn, Germany
 }%
 \author{J.A.~OLLER}
\address{ Departamento de F\'{\i}sica, Universidad de Murcia, 30071 Murcia, Spain
 }%
 \author{E.~OSET, M.~J.~VICENTE-VACAS}
\address{ Departamento de F\'{\i}sica Te\'orica and IFIC,
Centro Mixto Universidad de Valencia-CSIC,
 Apartado 22085,
 E-46071 Valencia, Spain }%



\maketitle

\abstracts{
We report on the latests developments in the field
of baryonic resonances generated
from the meson-baryon interaction in coupled channels using a chiral
unitary approach. The collection of resonances found in different strangeness and isospin sectors can be classified into SU(3) multiplets. The $\Lambda(1405)$ emerges as containing the effect of two poles of the scattering amplitude and various reactions that might preferentially select one or the other pole are discussed.
}

\section{Introduction}

Establishing the nature of hadronic resonances is one of the
primary goals in the field of hadronic physics. The interest lies in
understanding whether they
behave as genuine three quark states or they are dynamically
generated through the iteration of appropriate non-polar terms of
the hadron-hadron interaction, not being preexistent states
that remain in the large $N_c$ limit where the multiple scattering is suppressed.
In the last decade, chiral
perturbation theory ($\chi$PT) has emerged as a powerful scheme to
describe low-energy meson-meson and meson-baryon dynamics. In recent years,
the introduction
of unitarity constraints has allowed the extension of the
chiral description to much higher energies and, in addition,
it has lead to the generation of many hadron resonances both in the
mesonic and the baryonic sectors.

The $\Lambda(1405)$ resonance is a clear example of a dynamically
 generated state appearing naturally from the multiple scattering
  of coupled meson-baryon channels with strangeness $S=-1$ [\refcite{Jones:1977yk}--\refcite{Oller:2000fj}].
Recently, the interest in studying its properties has been revived
by the observation in the chiral models that the nominal
$\Lambda(1405)$ is in fact built up from two poles of the T-matrix
in the complex plane [\refcite{Oller:2000fj}--\refcite{Garcia-Recio:2002td}] both
contributing to the invariant $\pi\Sigma$ mass distribution, as it
was the case within the cloudy bag model [\refcite{Fink:1990uk}]. The
fact that these two poles have different widths and partial decay
widths into $\pi \Sigma$ and ${\bar K} N$ states opens the possibility
that they might be experimentally observed in hadronic or
electromagnetic reactions.

The unitary chiral dynamical models have been extended by various groups 
[\refcite{Oller:2000fj},\refcite{Garcia-Recio:2002td},\refcite{NPOR00}--\refcite{LK02}],
covering an energy range of about 1.4--1.7 GeV and
giving rise to a
series of resonant states in 
all isospin and strangeness sectors. All these observations have finally
merged into the classification of the dynamical generated baryon
resonances into SU(3) multiplets
[\refcite{JOORM03}], as seen also in Ref.~[\refcite{GLN03}].

In this contribution we present a summary of our latest
developments in the field of baryon resonances generated from
chiral unitary dynamics.

\section{Meson-baryon scattering model}

The search for dynamically generated resonances proceeds by first
constructing the meson-baryon coupled states from the octet of
ground state positive-parity baryons ($B$) and the octet of
pseudoscalar mesons
($\Phi$) for a given strangeness channel. Next, from the lowest
order lagrangian
\begin{equation}
L_1^{(B)} = \langle \bar{B} i \gamma^{\mu} \frac{1}{4 f^2} [(\Phi
\partial_{\mu} \Phi - \partial_{\mu} \Phi \Phi) B - B (\Phi
\partial_{\mu} \Phi - \partial_{\mu} \Phi \Phi)] \rangle
\end{equation}
one derives the driving kernel in s-wave
\begin{equation}
{V_{i j}} = - {C_{i j} \frac{1}{4 f^2}(2\sqrt{s} - M_{i}-M_{j})}
{\left(\frac{M_{i}+E_i}{2M_{i}}\right)^{1/2}
\left(\frac{M_{j}+E_j}{2M_{j}} \right)^{1/2}} \ , \label{eq:v}
\end{equation}
where the constants $C_{ij}$ are SU(3) coefficients encoded in the
chiral lagragian and $f$ is the meson decay constant, which we
take to have an average value of $f=1.123 f_\pi$, where $f_\pi=92.4$ MeV is
the pion decay constant. While at lowest order in the chiral
expansion all the baryon mases are equal to the chiral mass $M_0$,
the physical masses are used in Eq.~(\ref{eq:v}) as done in
Refs.~[\refcite{OR98},\refcite{ORB02}]. We recall that, in addition to the
Weinberg-Tomozawa or seagull term of Eq.~(\ref{eq:v}), one also
has at the same order of the chiral expansion the direct and
exchange diagrams considered in Ref.~[\refcite{Oller:2000fj}]. Their
contribution increases with energy and represents around 20\% of
that from the seagull term at $\sqrt{s}\simeq 1.5$ GeV.

The scattering matrix amplitudes between the various meson-baryon
states are obtained by solving the coupled channel equation
\begin{equation}
T_{ij} = V_{ij} + V_{il} G_{l} T_{lj} \ , \label{eq:BS} 
\end{equation}
where $i,j,l$ are channel indices and the $V_{il}$ and $T_{lj}$
amplitudes are taken on-shell. This is a particular case of the
N/D unitarization method when the unphysical cuts are ignored
[\refcite{OO99},\refcite{OM00}]. Under these conditions the diagonal matrix
$G_l$ is simply  built from the convolution of a meson and a
baryon propagator and can be regularized either by a cut-off
($q_{\rm max}^l$), as in Ref.~[\refcite{OR98}], or alternatively
by dimensional regularization depending on a subtraction constant
($a_l$) coming from a subtracted dispersion relation
[\refcite{Oller:2000fj},\refcite{ORB02}].

\section{Strangeness $S=-1$}

\begin{figure}
   \begin{center}
   \epsfxsize=7.5cm
   \epsfbox{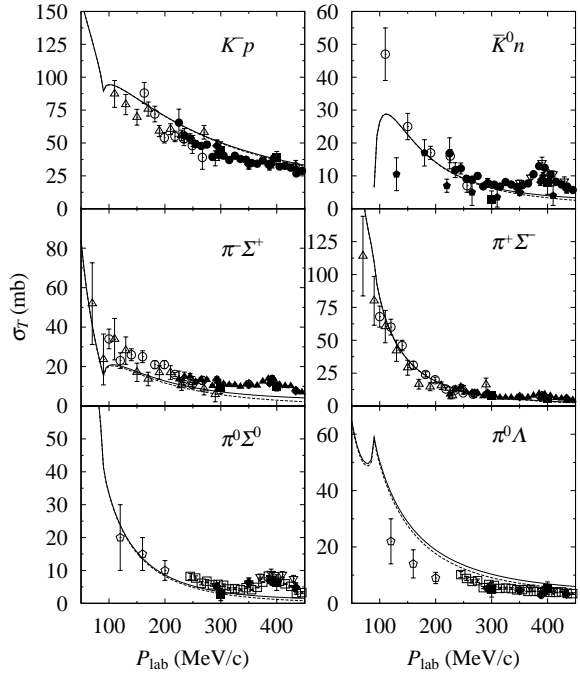}
   \end{center}
   \caption{Total cross sections of the $K^{-}p$ elastic and
   inelastic scatterings. The solid line denotes our results
    including both s-wave\ and p-wave.
   The dashed line shows our results without the p-wave amplitudes.
   The data are taken from Ref.~[\protect\refcite{exp}].
    \label{fig:cross}}
\end{figure}

\begin{figure}[ht]
\centerline{
     \includegraphics[width=2 in,angle=-90]{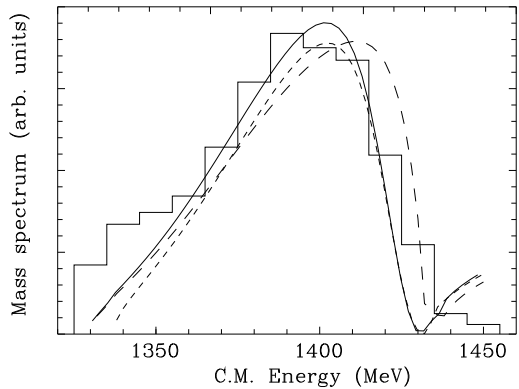}}
\caption{The $\pi\Sigma$ invariant mass distribution around the
$\Lambda(1405)$ resonance. Results in particle basis (solid line),
isospin basis (short-dashed line) or omitting the $\eta\Lambda$,
$\eta\Sigma^0$ channels. Experimental histogram taken from
Ref.~[\protect\refcite{He85}].\label{fig:reson}}
\end{figure}

In the case of $K^- p$ scattering, we consider the complete basis
of meson-baryon states, namely $K^- p$, $\bar{K}^0 n$,
$\pi\Lambda$, $\eta \Lambda$, $\eta\Sigma^0$, $\pi^+\Sigma^-$,
$\pi^-\Sigma^+$,$\pi^0\Sigma^0$,$K^+\Xi^-$ and $K^0 \Xi^0$, thus
preserving SU(3) symmetry in the limit of equal baryon and meson
masses. Taking a cut-off of 630 MeV, the scattering observables,
threshold branching ratios and properties of the $\Lambda(1405)$
resonance were well reproduced [\refcite{OR98},\refcite{jidopwave}],
as shown in Table \ref{tab:ratios}
and Figs.~\ref{fig:cross} and \ref{fig:reson}.
The inclusion of the $\eta\Lambda,\eta\Sigma$ channels was found crucial to obtain a good agreement with experimental data in terms of the lowest order chiral lagrangian.

\begin{table}[tbp]
\tbl{Branching ratios
 at $K^- p$ threshold.
Experimental values taken from Refs.~[\protect\refcite{To71,No78}].} 
{
\renewcommand{\tabcolsep}{0.5pc} 
\renewcommand{\arraystretch}{1.1} 
\begin{tabular}{|l|c|c|}
\hline
{} &{} &{} \\[-1.5ex]
{Ratio} & Exp. & Model \\[1ex]
\hline
{} &{} &{} \\[-1.5ex]
$\gamma= {\frac{\Gamma (K^- p \rightarrow \pi^+ \Sigma^-)} {\Gamma
(K^- p \rightarrow \pi^- \Sigma^+)}}$ & $2.36\pm 0.04$ & $2.32$ \\[1ex]
${ R_c} = { \frac{\Gamma (K^- p \rightarrow \hbox{\scriptsize charged
particles)}} {\Gamma (K^- p \rightarrow \hbox{\scriptsize all)}}}$ & $0.664\pm
0.011$ & $0.627$ \\[1ex]
${R_n} = {\frac{\Gamma (K^- p \rightarrow \pi^0 \Lambda)} {\Gamma
(K^- p \rightarrow \hbox{\scriptsize all neutral states})}}$ & $0.189\pm
0.015$ & $0.213$ \\[1ex]
\hline
\end{tabular}
\label{tab:ratios} }
\end{table}

\begin{table}[tbp]
\tbl{ Pole positions and couplings to meson-baryon states of
the dynamically generated resonances in the $S=-1$ sector
[\protect\refcite{ORB02}].}
{\label{tab:table1}
\newcommand{\m}{\hphantom{$-$}}
\newcommand{\cc}[1]{\multicolumn{1}{c}{#1}}
\renewcommand{\tabcolsep}{0.5pc} 
\renewcommand{\arraystretch}{1.1} 
\begin{tabular}{lcccccc}
\hline
 &  $z_R$ (MeV) &  &
$\mid g_{\pi\Sigma} \mid^2$ & $\mid g_{{\bar K}N} \mid^2$ & $\mid
g_{\eta\Lambda} \mid^2$ &
$\mid g_{K\Xi} \mid^2$ \\
\hline
$\Lambda(1405)$  & 1390$-$i66 & & 8.4 & 4.5 & 0.59 & 0.38 \\
 & 1426$-$i16  & & 2.3 & 7.4 & 2.0 & 0.12 \\
\hline
$\Lambda(1670)$ & 1680$-$i20 & & 0.01 & 0.61 & 1.1 & 12 \\
\hline
 &  & $\mid g_{\pi\Lambda} \mid^2$ 
& $\mid g_{\pi\Sigma} \mid^2$ & $\mid g_{{\bar K}N} \mid^2$ & $\mid g_{\eta\Sigma} \mid^2$ &
$\mid g_{K\Xi} \mid^2$ \\
\hline
$\Sigma(1620)$ & 1579$-$i274 & 4.2 & 7.2 & 2.6 & 3.5 & 12 \\
\hline
\end{tabular}\\[2pt]}
\end{table}

Our model extrapolated smoothly to high energies [\refcite{ORB02}] by
using the dimensional regularization scheme. The subtraction
constants resulted to have a ``natural" size [\refcite{Oller:2000fj}] which
permits qualifying the generated resonances as being dynamical.
While the $I=0$ components of the $\bar{K}N\to{\bar K}N$ and
$\bar{K}N\to\pi\Sigma$ amplitudes displayed a clear signal from
the $\Lambda(1670)$ resonance, the $I=1$ amplitudes showed to be
smooth and featureless without any trace of resonant behavior, in
line with the experimental observation. In Table~\ref{tab:table1} we display the value of the
poles of the scattering amplitude in the second Riemann
sheet, $z_R=M_R - {\rm i}\Gamma/2$, together with the
corresponding couplings to the various meson-baryon states,
obtained from identifying the amplitudes $T_{ij}$ with $g_i
g_j/(z-z_R)$ in the limit $z\to z_R$.
Two poles define the $\Lambda(1405)$ resonance. The pole at lower energy is
wider and couples mostly to $\pi\Sigma$ states, while that at higher energy is narrower and couples mostly to ${\bar K}N$ states. The consequences of this two-pole nature of the $\Lambda(1405)$ are discussed in detail in Refs.~[\refcite{JOORM03},\refcite{jidohere}]. 
We also find poles corresponding to the
$\Lambda(1670)$ and $\Sigma(1620)$ resonances. The large coupling
of the $\Lambda(1670)$ to $K\Xi$ states allows one to identify
this resonance as a ``quasibound" $K\Xi$ state. The large width
associated to the $\Sigma(1620)$ resonance, rated as 1-star by the
Particle Data Group (PDG) [\refcite{Hagiwara:2002fs}], explains why there is no trace of this state in the
scattering amplitudes.

\section{Strangeness $S=-2$}

\begin{figure}[hbt]
\begin{center}
\includegraphics[width=5cm,angle=-90]{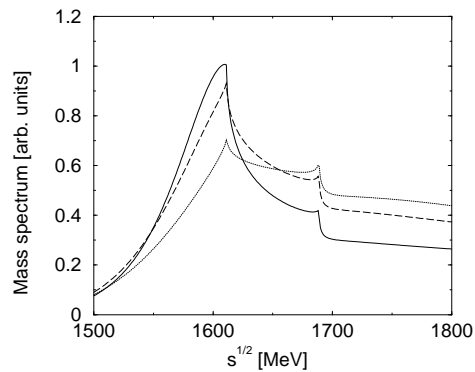}
\caption{The $\pi\Xi$ invariant mass distribution as a function of
the center-of-mass energy, for several sets of subtraction
constants. Solid line: $a_{\pi \Xi}=-3.1$ and
$a_{\bar{K}\Lambda}=-1.0$; Dashed line: $a_{\pi \Xi}=-2.5$ and
$a_{\bar{K}\Lambda}=-1.6$; Dotted line: $a_{\pi \Xi}=-2.0$ and
$a_{\bar{K}\Lambda}=-2.0$. The value of the two other subtraction
constants, $a_{{\bar K}\Sigma}$ and $a_{\eta\Xi}$, is fixed to
$-2.0$ in all curves.} \label{fig:masspixi}
\end{center}
\end{figure}

The unitariy chiral meson-baryon approach has also been extended
to the $S=-2$ sector [\refcite{ROB02}] to investigate the nature of the
lowest possible s-wave $\Xi$ states, the $\Xi(1620)$ and
$\Xi(1690)$, rated 1- and 3-star, respectively, and quoted with
unknown spin and parity by the PDG [\refcite{Hagiwara:2002fs}]. Allowing the subtraction
constants to vary around a natural size of $-2$, a pole is found
at $z_R=1605-{\rm i}66$, the real part showing a strong stability
against the change of parameters. The imaginary part would
apparently give a too large width of 132 MeV compared to the
experimental ones reported to be of 50 MeV or less. However, due a
threshold effect, the actual $\pi\Xi$ invariant mass distribution,
displayed in Fig.~\ref{fig:masspixi}, shows a much narrower width
and resembles the peaks observed experimentally.

The couplings obtained are $\mid g_{\pi\Xi}\mid^2=5.9$, $\mid
g_{{\bar K}\Lambda}\mid^2=7.0$, $\mid g_{{\bar
K}\Sigma}\mid^2=0.93$ and $\mid g_{\eta\Xi}\mid^2=0.23$. The
particular large values for final $\pi\Xi$ and ${\bar K}\Lambda$
states rule out identifying this resonance with the $\Xi(1690)$,
which is found to decay predominantly to ${\bar K}\Sigma$ states.
Therefore, the dynamically generated $S=-2$ state can be safely
identified with the $\Xi(1620)$ resonance and this also allows us
to assign the values $J^P=1/2^-$ to its unmeasured spin and
parity. The model of Ref.~[\refcite{GLN03}] finds this state at
$z_R=1565-{\rm i}124$, together with another pole at $z_R=1663-{\rm i}2$, identified with the $\Xi(1690)$ because of its strong coupling
to ${\bar K}\Sigma$ states.

\section{Strangeness $S=0$}

For completeness, we briefly mention here the work done in the $S=0$ sector 
[\refcite{NPOR00},\refcite{IOV02}] where the $N(1535)$ was generated dynamically within
the same approach. In order to reproduce the phase shifts and inelasticities,
four subtraction constants were adjusted to the data leading to
a $N(1535)$ state with a total decay width of $\Gamma   
\simeq 110$ MeV, divided into $\Gamma_\pi\simeq 43$ MeV and 
$\Gamma_\eta\simeq 67$ MeV,
compatible with present data within errors.
The dynamical $N(1535)$ is found to have strong couplings to the $K\Sigma$ 
and $\eta N$ final states.

\section{SU(3) multiplets of resonant states}

\begin{figure}
  \epsfxsize=10cm
  \centerline{\epsfbox{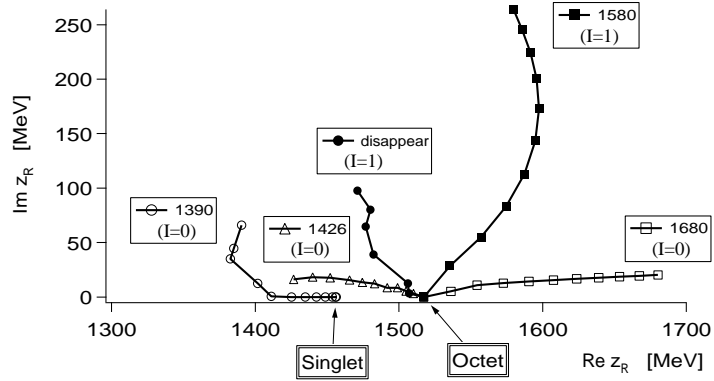}}
  \caption{Trajectories of the poles in the scattering amplitudes obtained by
  changing the SU(3) breaking parameter $x$ gradually. At the SU(3) symmetric
  limit ($x=0$),
   only two poles appear, one is for the singlet and the other (two-times     degenerate) for the octets.
  The symbols correspond to the step size $\delta x =0.1$. The results are from
  Ref.~[\protect\refcite{JOORM03}].
  \label{fig:tracepole}}
\end{figure}

The SU(3) symmetry encoded in the chiral lagrangian permits
classifying all these resonances into SU(3) multiplets. 
We first recall that the meson-baryon states built from the
octet of pseudoscalar mesons and the octet of ground state baryons
can be classified into the irreducible representations:
\begin{equation}
8\otimes 8 = 1\oplus 8_s \oplus 8_a \oplus 10 \oplus \bar{10}
\oplus 27
\end{equation}
Taking a common meson mass and a common baryon mass,
the lowest-order meson-baryon chiral lagrangian is exactly
SU(3) invariant. 
If, in addition, all the subtraction constants
$a_l$ are equal to a common value, the scattering problem decouples
into each of the SU(3) sectors. Using SU(3) Clebsh-Gordan coefficients, 
the matrix elements of the transition potential $V$ in a
basis of SU(3) states are
\begin{equation}
V_{\alpha\beta} \propto -\frac{1}{4f^2}\sum_{i,j} \langle i,\alpha\rangle
C_{ij} \langle j,\beta\rangle =\frac{1}{4f^2} {\rm diag}
(-6,-3,-3,0,0,2) \ ,
\end{equation}
taking the following order for the irreducible representations:
$1$,$8_s$,$8_a$,$10$,$\bar{10}$ and $27$.
The attraction in the singlet and the two octet channels gives rise to
bound states in the
unitarized amplitude, with the two octet poles being degenerate [\refcite{JOORM03}]. By breaking the
SU(3) symmetry gradually, allowing the masses and subtraction constants to evolve to their physical values, the degeneracy is lost and the poles move along trajectories in the complex plane as shown in Fig.~\ref{fig:tracepole}, which collects 
the behavior of the $S=-1$ states. As discussed further in  
Refs.~[\refcite{JOORM03},\refcite{jidohere}], two poles in the $I=0$ sector appear very close in energy and they will manifest themselves as
a single resonance, the $\Lambda(1405)$, in invariant $\pi\Sigma$ mass distributions.

\section{The two-pole nature of the $\Lambda(1405)$}

The $\Lambda(1405)$ is seen through the invariant mass distributions of $\pi\Sigma$ states given by
\begin{equation}
    \frac{d\sigma}{d M_I}=|\sum_{i}C_it_{i\to\pi\Sigma}|^2p_{CM}
    \label{eq:mdist2}
\end{equation}
with $i$ standing for any of the coupled channels ($\bar{K}N$, $\pi\Sigma$, $\eta\Lambda$, $K\Xi$) and $C_i$ being coefficients that determine the strength for the excitation of 
channel $i$, which eventually evolves into a $\pi\Sigma$ state through the multiple scattering. As the two $\Lambda(1405)$ poles couple differently to $\pi\Sigma$ and ${\bar K}N$ states, the amplitudes 
$t_{\pi\Sigma \to \pi\Sigma}$, $t_{{\bar K}N \to \pi\Sigma}$ are dominated by one or the other pole, respectively, thus making the invariant mass distribution sensitive to the coefficients $C_{\pi\Sigma}$, $C_{{\bar K}N}$, {\it i.e.} to the reaction used to generate the $\Lambda(1405)$.

An interesting example is found in the radiative production reaction
$K^- p \to \gamma \Lambda(1405)$. In order to access the subthreshold region, the photon must 
be radiated from the initial $K^- p$ state, ensuring that the $\Lambda(1405)$ resonance is
initiated from $K^- p$ states, hence selecting the pole that
couples more strongly to $\bar{K}N$ which is narrower and appears
at a higher energy. The calculated
invariant mass $\pi\Sigma$ distribution [\refcite{NOTR99b}]
appears indeed displaced to higher energies ($\sim 1420$ MeV) and it is
narrower (35 MeV) than what one obtains from other  reactions.

The $\Lambda(1405)$ can also be produced from the reaction
$(\gamma,K^+)$ on protons, recently implemented at LEPS of
SPring8/RCNP [\refcite{ahn}]. In this case, the invariant mass
distribution of the final meson-baryon state obtained in
Ref.~[\refcite{NOTR99a}] shows a width of around 50 MeV.
Due to the particular
isospin decomposition of the $\pi\Sigma$ states, the $\pi^-\Sigma$
and $\pi^+\Sigma^-$ cross sections differ in the sign of the
interference between $I=0$ and $I=1$ amplitudes (omitting the
negligible $I=2$ contribution). This difference has been observed
in the experiment performed at SPring8/RCNP [\refcite{ahn}] and
provides some information on the $I=1$ amplitude.

\begin{figure}[tbp]
    \centering
    \includegraphics[width=11cm,clip]{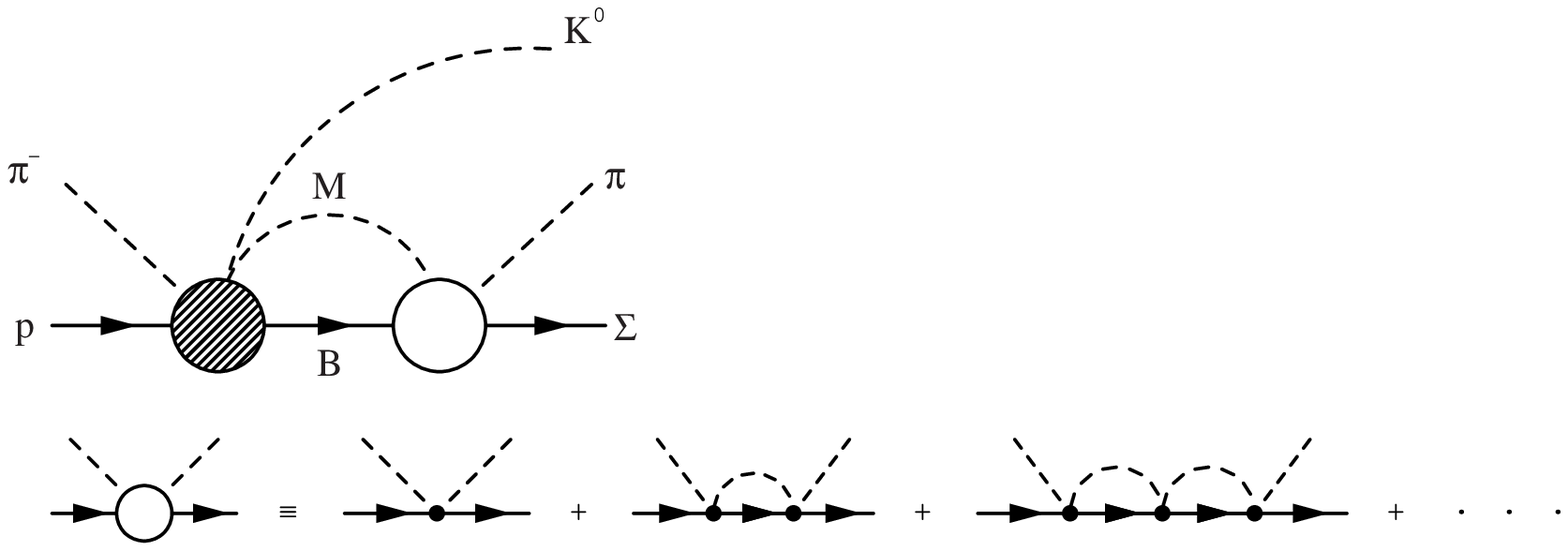}
    \includegraphics[width=7.5cm,clip]{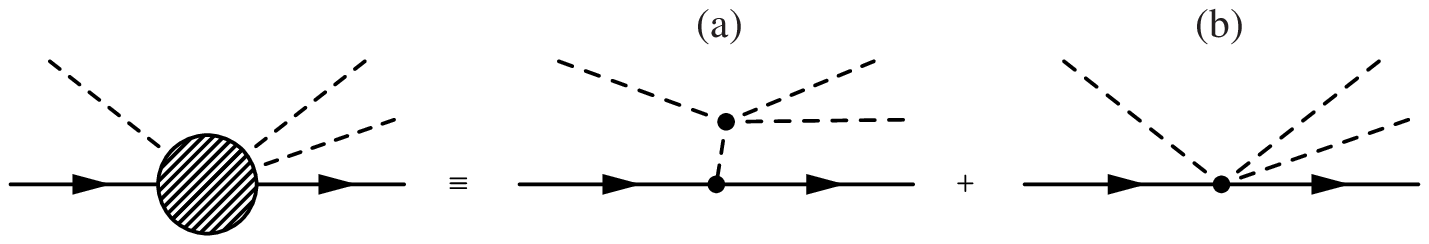}
    \caption{\label{fig:1}
    Diagrams entering the production of the $\Lambda(1405)$ in the
    reaction $\pi^- p\to K^0\Lambda(1405) \to K^0\pi\Sigma$.
    }
\end{figure}

The dynamics
that goes into the $\pi^-p\to K^0\pi\Sigma$ reaction, from which the experimental data
of the $\Lambda(1405)$ resonance have been extracted [\refcite{Thomas:1973uh}], has recently been investigated  [\refcite{HHORV03}].
As shown in Fig.~\ref{fig:1}, the process is separated into a part
which involves tree level $\pi^-p\to K^0MB$ amplitudes (hatched blob), and a
second part which involves the final state interaction $MB\to
\pi\Sigma$.
The initial process is described following the model
for the $\pi N\to \pi\pi N$ reaction close to
threshold, which contains a pion pole term and a contact term, both
of them calculated from the chiral lagrangians.

\begin{figure}[tbp]
    \centering
    \includegraphics[width=11cm,clip]{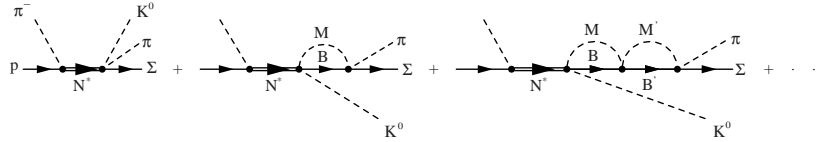}
    \caption{\label{fig:6}
    Resonant mechanisms for $\Lambda(1405)$ production in the
    $\pi^-p\to K^0\pi\Sigma$ reaction.}
\end{figure}

Since in this reaction $\sqrt{s}\sim 1900$ MeV, one must also consider
resonance excitation in the $\pi N$
collision leading to the decay of the resonance in $MMB$, as seen
in Fig.~\ref{fig:6}. 
We select the $N^*(1710)$ since,
in the energy region of interest, it is the only $S=0$ $P_{11}$
resonance with the same quantum
numbers of the nucleon having a
very large branching ratio to
$\pi\pi N$ (40-90\%) [\refcite{Hagiwara:2002fs}].

\begin{figure}[tbp]
    \centering
    \includegraphics[width=7cm,clip]{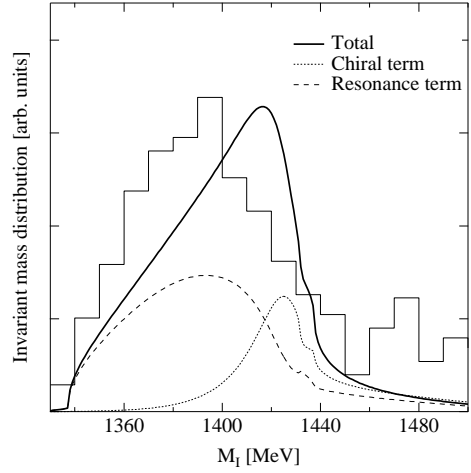}
    \caption{\label{fig:155mdist100}
    Invariant mass distribution of $\pi\Sigma$ obtained by averaging
    $\pi^+\Sigma^-$ and $\pi^-\Sigma^+$.
    The histogram shows the
    experimental data taken from Ref.~[\protect\refcite{Thomas:1973uh}].
     Resonance parameters: $M_R = 1740$, $\Gamma_{N^*}=200$ MeV,
    $\Gamma_{\pi N}=30$ MeV     and
    $\Gamma_{\pi\pi N}=120$ MeV.}
\end{figure}%

The contribution of the chiral and resonant mechanisms to the invariant mass distribution 
are shown in Fig.~\ref{fig:155mdist100} by the dotted and dashed lines, respectively. Both contributions are of similar size and their coherent sum
(solid line) produces a distribution more in agreement with the experimental histogram.
The chiral
tree amplitude $\pi^-p\to
K^0M_iB_i$ for the case $M_iB_i=\bar{K}N$ involves the
combinations $3F-D$ and $D+F$, which are large compared to the
$D-F$ combination that one finds for $M_iB_i\equiv\pi\Sigma$ (we
take $F=0.51$ and $D=0.75$). Therefore, the chiral distribution gives 
a larger weight to the $t_{\bar{K}N\to\pi\Sigma}$ amplitude, which
is dominated by the narrower pole at higher energy.
On the contrary, the
the $N^*\to B M_1M_2$
vertex in the resonant mechanism goes like 
the difference of energies of the outgoing mesons prior to final state interaction effects, which is practically zero for $N\bar{K} K^0$ and
300 MeV for $\Sigma \pi K^0 $.
 Therefore, the resonant contribution is strongly dominated
 by the $t_{\pi\Sigma\to\pi\Sigma}$ amplitude, which couples more
 strongly to the wider pole at lower energy.

Recently, the production of the $\Lambda(1405)$ through $K^*$ vector meson photoproduction, $\gamma p \to K^* \Lambda(1405) \to \pi K \pi \Sigma$, using linearly polarized photons has been studied [\refcite{HHVO04}].
Selecting the events in which the polarization of the incident photon and that of the produced $K^*$ are perpendicular, the mass distribution of the $\Lambda(1405)$  peaks at 1420 MeV, since in this case the process is dominated by $t$-channel K-meson exchange, hence selecting preferentially the
 $t_{{\bar K} N\to {\bar K}N}$ amplitude.

\section{Summary and Conclusions}

By implementing unitarity in the study of meson-baryon scattering using the lowest order chiral lagrangian, a series of resonant states have been dynamically generated in all strangeness and isospin sectors.

In the SU(3) limit, all these resonances belong to a singlet or to either of the two (degenerate) octets of dynamically generated poles of the SU(3) symmetric scattering amplitude. 

In the physical limit, there are two $I=0$ poles representing the $\Lambda(1405)$, the one at lower energy having a larger imaginary part than the one at higher energy. These poles couple differently to $\pi\Sigma$ and ${\bar K}N$ states and, as a consequence, the properties of the $\Lambda(1405)$ will depend on the particular reaction used to produce it. Various processes that might preferentially select the contribution from
one or the other pole have been discussed.

\section*{Acknowledgments}
This work is supported by DGICYT (Spain) projects BFM2000-1326,
BFM2002-01868 and FPA2002-03265, the EU network EURIDICE contract
HPRN-CT-2002-00311, and the Generalitat de Catalunya project
2001SGR00064.

%
%
%
%
%
%
%
%
%
%
%
%
%
%
%

\end{document}